\documentclass[aps,pra,11pt]{revtex4}  

\usepackage{amsmath}
\usepackage{mathrsfs}
\usepackage{amssymb}
\usepackage{graphicx,epstopdf,epsfig}
\usepackage{bm}
\usepackage{color}
\usepackage{times}
\usepackage{hyperref,url}
\usepackage{hyperref}
\usepackage{url}

\newcommand{\mr}[1]{{\mathrm{#1}}} 		
\newcommand{\ket}[1]{{\vert #1 \rangle}} 		

\definecolor{clr}{rgb}{0,0.6,0.6}

\begin{document}
\title{Sub-kHz excitation lasers for Quantum Information Processing with Rydberg atoms}

\author{R. Legaie}
\author{C. J. Picken}
\author{J. D. Pritchard}
\email{jonathan.pritchard@strath.ac.uk}
\affiliation{Department of Physics, Strathclyde University, Glasgow, G4 0NG, UK}

\date{\today}

\begin{abstract}
Quantum information processing using atomic qubits requires narrow linewidth lasers with long-term stability for high fidelity coherent manipulation of Rydberg states. In this paper, we report on the construction and characterization of three continuous-wave (CW) narrow linewidth lasers stabilized simultaneously to an ultra-high finesse Fabry-Perot cavity made of ultra-low expansion (ULE) glass, with a tunable offset-lock frequency. One laser operates at 852~nm while the two locked lasers at 1018~nm are frequency doubled to 509~nm for excitation of $^{133}$Cs atoms to Rydberg states. The optical beatnote at 509~nm is measured to be 260(5)~Hz. We present measurements of the offset between the atomic and cavity resonant frequencies using electromagnetically induced transparency (EIT) for high-resolution spectroscopy on a cold atom cloud. The long-term stability is determined from repeated spectra over a period of 20 days  yielding a linear frequency drift of $\sim1$~Hz/s.
\end{abstract}

\maketitle
\section{Introduction}


The field of quantum information processing (QIP) is an intensive research area. Its attractiveness lies in the possibility of speeding up classical problems and modelling complex quantum systems \cite{nielsen05}. Neutral atoms present an attractive candidate for scalable QIP \cite{saffman16,auger17} combining long coherence times of weakly interacting hyperfine ground states \cite{treutlein04} with strongly interacting Rydberg states \cite{saffman10} to create pair-wise entanglement \cite{wilk10,jau16}, perform deterministic quantum gates \cite{isenhower10,maller15} and even realize a quantum simulator for Ising models \cite{labuhn16}. 
 
Rydberg excitation is typically performed using two-photon excitation due to weak single photon matrix elements from the ground state \cite{johnson08} and inconvenient UV wavelengths. Using a resonant two-photon excitation, Rydberg electromagnetically induced transparency (EIT) \cite{mohapatra07} can be exploited for laser frequency stabilization \cite{abel09} as well as precision metrology of Rydberg state energies \cite{mack11,grimmel15} and lifetimes \cite{mack15}, dc electric fields \cite{tauschinsky10} and RF field sensors operating at both microwave \cite{sedlacek12,fan15} and THz \cite{wade17} frequency ranges. For dense cold atom samples, the strong atomic interactions can be mapped onto the optical field to create non-linearities at the single photon level \cite{pritchard13,firstenberg16,gorniaczyk16,tiarks16}.

For robust Rydberg excitation of atomic qubits for gate operations the two-photon excitation must be detuned from the intermediate excited state to avoid losses due to spontaneous emission. High fidelity gates also require narrow linewidth excitation lasers with excellent long term frequency stability \cite{cummins03}. These requirements can be met using lasers stabilized to a high-finesse optical cavity \cite{fox03,martin12}, exploiting techniques developed for lasers on state-of-the-art optical lattice clocks operating at fractional instabilities $\mr{< \,10^{-18}}$ \cite{bloom14,campbell17} requiring sub-Hz laser linewidths \cite{alnis08,cappellini15,hill16}. Using cavities made of ultra-low expansion (ULE) glass, enables minimization of the long term drifts to $\mr{< \, 0.1 \, Hz/s}$ \cite{alnis08}, rivaled only by the performance of cryogenically cooled single crystal silicon cavities \cite{hagemann14,matei17}.

Recently, details of cavity stabilized lasers systems for Rydberg excitation of K \cite{arias17}, Rb \cite{dehond17}, and Sr \cite{bridge16} have been presented, achieving typical linewidths around 1-10~kHz. In this paper, we describe the stabilization of three continuous-wave lasers to a ULE reference cavity for Rydberg excitation of Cs, offering sub-kHz linewidths for implementation of QIP Rydberg protocols. We determine a lock bandwidth of 1.1~MHz via observation of the in-loop power spectrum. High resolution EIT spectroscopy of cold Cs Rydberg states is used to calibrate the cavity mode frequencies with respect to Rydberg transitions and determine the cavity long term drift $ \mr{\sim  \,1 \, Hz/s} $.

\section{Laser sources}

\begin{figure}
\centering \includegraphics{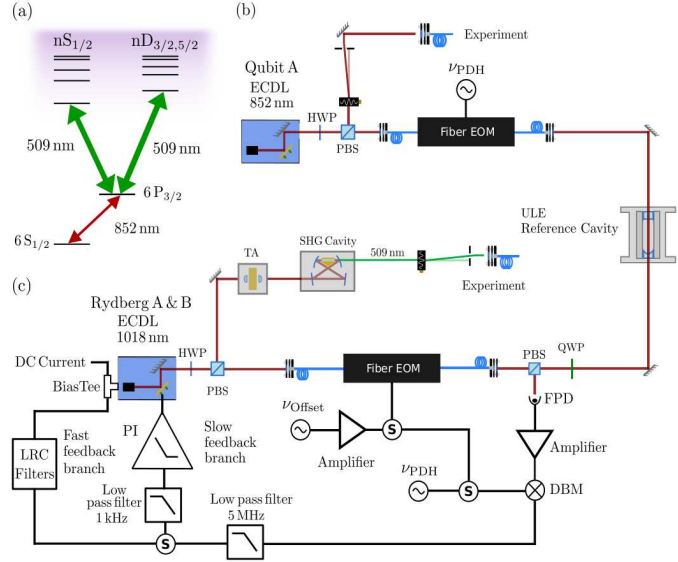}
\caption{(a) Schematic of the two-photon excitation of $^{133}$Cs to $nS_{1/2}$ or $nD_{3/2,5/2}$ Rydberg state via $6P_{3/2}$ state. (b) Laser setup for Qubit~A at 852~nm driving transition from $6S_{1/2}\rightarrow6P_{3/2}$ (c) Locking electronics for frequency doubled Rydberg lasers A \& B driving the second stage transition at 509~nm. Key~: PBS = Polarizing beam splitter, HWP =~Half wave plate, QWP = Quarter wave plate, FPD = Fast photodiode, DBM = Double balanced mixer, S = Splitter, PI = Proportional Integral, TA~=~Tapered amplifier, AOM =  Acousto-optic modulator, EOM = Electro-optic modulator.}
\label{fig1}
\end{figure}

We perform Rydberg excitation via the Cs $D_2$ line using transitions from $6s~^2S_{1/2}\rightarrow6p~^{2}P_{3/2}\rightarrow n\ell~^{2}L_j$ as shown in Fig.~\ref{fig1}(a). The first step of the excitation from $6s~^2S_{1/2}\rightarrow6p~^{2}P_{3/2}$ is performed using an extended cavity diode laser (ECDL) at 852~nm, as shown in Fig.~\ref{fig1}(b). For the second step, light at 509~nm is required which is generated from a pair of homebuilt second harmonic generation (SHG) systems, Rydberg A \& B, which double light from a pair of master lasers operating at 1018~nm as shown in Fig.~\ref{fig1}(c). 

Our ECDL design uses laser diodes mounted in a solid aluminium body with frequency control achieved using holographic gratings operated in a Littrow configuration \cite{arnold98}. The grating mount is doubly hinged to provide decoupling of horizontal and vertical adjustment. Coarse frequency tuning of the grating angle is performed using a precision 170~TPI screw, with fine tuning achieved using a piezoelectric actuator (Thorlabs AE0505D08F). All the lasers are temperature stabilized to $< 0.1^\circ$C using a commercial Arroyo 5240 PID controller driving a Peltier cooler and placed in a perspex box with an AR coated window to provide thermal and mechanical insulation.

The first step laser, Qubit A, uses a standard Fabry-Perot laser diode (Thorlabs L852P150) with $30~\%$ optical feedback from a grating with 1800~lines/mm (Thorlabs GH13-18V) giving a mode-hop free tuning range of 5~GHz. After the isolator, light passes through a noise-eater AOM to eliminate intensity noise followed by a double-pass 80~MHz AOM for control of frequency and intensity before being delivered to the experiment via a polarization-maintaining (PM) single mode optical fiber.

The master lasers for Rydberg A \& B, are generated from AR coated infrared laser diodes at 1018nm (Toptica LD-1020-0400-2 and Sacher SAL-1030-060 respectively) combined with a 1200~lines/mm visible grating (Thorlabs GH13-12V) tunable from 1010-1025~nm with no realignment of the vertical feedback. High power is achieved using a Tapered-Amplifier (TA) (M2K TA-1010-2000-DHP) for each laser. We obtain 1.36 W output TA power for 22~mW seed power at a TA input current of 4A. 

To generate light at 509~nm, the Rydberg lasers are doubled via cavity-enhanced second harmonic generation (SHG) \cite{hemmerich90} using an AR coated quasi-phase matched periodically poled KTP crystal from Raicol Crystal Ltd. with dimensions of 1x2x20 mm and a poling period of $\Lambda$ = 7.725 $\mu$m. The non-linear crystal is placed into a brass heating block at the centre of a symmetrical bow-tie cavity, with temperature control provided using a Peltier device allowing temperature tuning from 15-95$^\circ$C. This corresponds to peak doubling efficiencies from 1020-1015~nm to enable excitation of Rydberg states with principal quantum number from $n\ge45$ to ionization. The bow-tie cavity consists of two concave mirrors with -30~mm radius of curvature and two plane mirrors, designed to achieve the optimal Boyd-Kleinmann waist of $25~\mu$m in the crystal \cite{boyd68}. The cavity has a 1.25~GHz FSR and a finesse of $\mathcal{F}\sim50$, to which we achieve 95~\% modematching to the light from the TA. 

The cavity length is stabilized to give peak SHG output power using the H\"ansch-Couillaud technique \cite{hansch80}, with a single proportional-integral (PI) servo feeding back to a ring piezo attached to one of the planar cavity mirrors. Following optimization of the alignment and crystal temperature, we typically obtain 370 mW at 509~nm for 650 mW infrared power, achieving  $\mr{\sim \, 57 \, \%}$ conversion efficiency at a TA current of 3.0~A. At higher infrared input powers ($>$ 1W), we observe power clamping arising from a competing $\chi$$^{(2)}$~-~non~linearity whereby the intracavity second harmonic beam acts as a pump for non-degenerate optical parametric oscillation \cite{white97}. This can be overcome using a lower input coupler reflectivity  \cite{ricciardi10} or by adjusting the cavity geometry to increase the waist in the crystal \cite{targat05}. Finally, the green light is then sent through an 80 MHz AOM  to provide intensity control and coupled into a single mode PM fiber, leading to $\sim160$~mW available at the cold atoms. AOM frequencies for all lasers are derived from a common direct digital synthesis (DDS) evaluation board (AD9959) device to provide controllable relative phase between the lasers.

\section{Laser frequency stabilization}
Laser stabilization is performed by locking each of the master lasers above to a stable high finesse reference cavity using the Pound-Drever-Hall (PDH) technique \cite{drever83, black01}. The reference cavity is a 10~cm cylindrical cavity from Advanced Thin Films in a plano-concave configuration with -50~cm radius of curvature. Both the mirrors and the spacer are made of ULE glass, making it possible to minimize thermal expansion of the cavity due to a zero-crossing of the linear coefficient of thermal expansion (CTE) \cite{fox09} providing long-term frequency stability. The mirror substrates have a dual-wavelength coating to provide high finesse at both 852~nm and 1018~nm simultaneously.

\begin{figure}
\centering \includegraphics{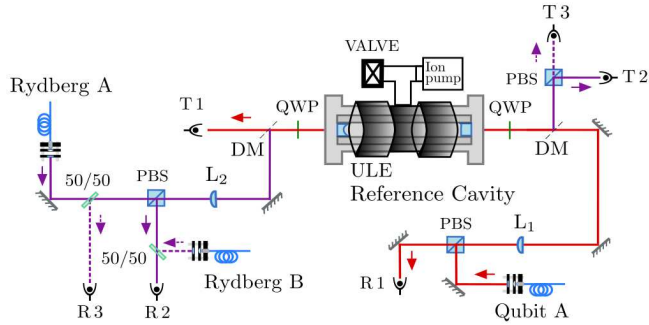}
\caption{Cavity locking setup used to lock two Rydberg lasers A \& B (1018~nm) and the Qubit A laser (852~nm) to an ULE reference optical cavity with a fast Pound-Drever-Hall lock. Key~: L = Cavity mode-matching lens, DM = Dichroic mirror, PBS = Polarizing beam splitter, QWP = Quarter wave plate, R~=~High-bandwidth photodiode, T = High-gain photodiode.}
\label{fig2}
\end{figure}

The optical reference cavity is housed in an evacuated vacuum chamber to minimize fluctuations in cavity frequency due to changes in the refractive index. Using the 3-l/s ion pump on the cavity chamber, we maintain a pressure of $3\times10^{-7}$~mbar. The cavity is mounted horizontally on a pair of Viton$^{\textregistered}$ O-rings placed at the Airy points \cite{chen06} to reduce sensitivity to vibration. The vacuum chamber is externally temperature-stabilized using a PI temperature controller that drives a current through heater tapes connected in parallel and wound around the cavity vacuum chamber, itself surrounded by 1 cm thick foam insulation. For passive vibration isolation, the cavity is mounted on a 60 x 60 cm$^{2}$ breadboard that rests on a layer of Sorbothane on a floating optical table. Further improvements could be achieved using a radiation shield around the cavity in vacuum or enclosing the cavity setup within a second stage of temperature control \cite{alnis08}.

The optical setup for laser locking is shown in Fig.~\ref{fig2}. We pick off $\sim1$~mW of light from each ECDL and couple the light into fiber-coupled electro-optic phase modulators (EOM) (Jenoptik models PM-830  and  PM-1064) for generating sidebands on the light. The fiber EOMs provide broadband phase modulation (up to 5~GHz) whilst minimizing residual amplitude modulation (RAM) and cleaning the mode shape and the polarization incident on the cavity. Due to the high insertion loss of the modulators, this leaves around 200~$\mu$W available for laser stabilization which is mode-matched into the cavity using lenses $f_{1,2}$ to maximize coupling to the TEM$_{00}$ cavity mode. Light from the two Rydberg master lasers (A $\&$ B) are combined on a polarizing beam splitter (PBS) with orthogonal polarization and coupled into the cavity using a dichroic mirror (DM) to separate the incident 1018~nm light from the transmitted 852~nm light from the cavity. To isolate the reflected signal for cavity locking, 50:50 beam splitters are placed in each beam path prior to the PBS allowing independent detection photodiodes to be used. Similarly, after the cavity a second DM and PBS are used to separate the transmission signals of the two Rydberg lasers. Finally, the transmitted and reflected signals from the cavity are monitored via a high-gain and a high-bandwidth home-built photodiodes respectively. The high-gain photodiode has been optimized for ring-down measurements on the ULE cavity, with a total gain of $2.5\times10^{6}$~V/A and a measured roll-off frequency of $F_{-3\mr{dB}}=890$~kHz. The high-bandwidth photodiode uses a fast photodetector (Hamamatsu S5971) with a single gain stage to provide a bandwidth of 25~MHz and a gain of $5\times10^{4}$~V/A. 

Electronics for the PDH lock are shown schematically in Fig.~\ref{fig1}(c). Each EOM is driven by a low frequency signal ($\nu_\mathrm{PDH}$) at +10~dBm to generate 1$^\mr{st}$-order sidebands with 10~\% amplitude (phase modulation index $\delta=0.7$~rad). To minimize cross-talk between lasers, frequencies of $\nu_\mathrm{PDH}=8.4,10,11.7$~MHz are chosen, ensuring any interference effects occur at beat frequencies above the servo-bandwidth. The PDH error signal is obtained by first amplifying the signal from the reflection photodiode using a low-noise amplifier and then demodulating at $\nu_\mathrm{PDH}$ on a mixer followed by a low-pass filter at 5~MHz. After the filter, the error signal is split into two simultaneous feedback branches to the laser. The first provides fast feedback directly to the diode laser current using a resistor for direct feedback in parallel with passive LRC phase advance and phase-delay filters as detailed in Ref.~\cite{fox03}. This is combined with the laser drive current using a bias-tee, and component values optimized to maximize the achievable servo bandwidth to 1.1~MHz which are measured by recording the in-loop photodiode signal using an rf spectrum analyzer as shown in Fig.~\ref{fig4} (a). A second, low frequency PI servo loop (DC-300~Hz) provides feedback to the laser piezo to ensure the laser remains locked to the cavity peak.

As the cavity resonances are not necessarily commensurate with frequencies required for Rydberg excitation, we employ the "electronic sideband" technique \cite{thorpe08, gregory15} to provide a continuously tunable offset from the cavity modes of the two Rydberg lasers. A second frequency, $\nu_\mr{Offset}$, is amplified to +25~dBm and combined on a splitter with the low frequency PDH signal, $\nu_\mr{PDH}$, to drive the EOM with dual frequencies resulting in large first order sidebands at $\pm\nu_\mr{Offset}$, each of which has secondary PDH sidebands to enable locking. Through choice of phase in the PDH error signal, the laser can be locked to either the +1 or -1 sideband to achieve a frequency shift of $\mp\nu_\mr{Offset}$ on the master laser with respect to the cavity. The offset frequency is derived from a DDS (AD9910) operating from 0.1 Hz to 460 MHz, which gives a tuning range of $\pm \, 920$~MHz after doubling.

\section{System Performance}

\begin{figure}[t!]
\centering \includegraphics{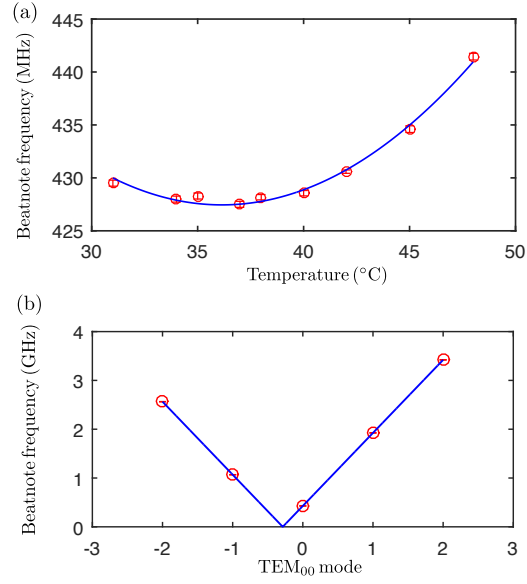}
\caption{(a) Determination of the zero-expansion temperature $\mr{T_{c}}$~=~36.1~$\mr{\pm}$~0.1~$^\circ$C with a parabolic fit. Operating at $\mr{T_{c}}$  gives a second order frequency sensitivity to temperature change. (b)~Free Spectral Range determination via the beatnote between Qubit A and a laser stabilized to the $F =  4$ to $F' = 5$ transition.}
\label{fig3}
\end{figure}

To evaluate the performance of the laser stabilization system, we first characterize the ULE cavity using an optical beatnote between the 852~nm Qubit A laser and a second independent laser stabilized to the $6s_{1/2}~F=4\rightarrow 6p_{3/2}~F'=5$ cooling transition using polarization spectroscopy \cite{pearman02}. Recording the beatnote frequency as a function of cavity temperature, as shown in Fig.~\ref{fig3}(a), results in a quadratic dependence due to the vanishing first-order coefficient of thermal expansion of the ULE cavity spacer \cite{alnis08}. Due to the large thermalization time-constant of the cavity (measured to be approximately 12~hours), each data point has been taken after a minimum period of 18 hours following a change in temperature and up to 48 hours later. From the data we extract the temperature of the zero-CTE crossing as $T_{c}=36.1\pm0.1^\circ$C. Following temperature stabilization of the cavity length at $T_c$, the free-spectral range is measured by locking the laser to adjacent longitudinal cavity mode, as shown in Fig.~\ref{fig3}(b). These data give $\nu_\mr{FSR}=1.49637(2)$~GHz, corresponding to a cavity length of $L=10.0173(1)$~cm.

Cavity finesse is measured using cavity ring-down, where an AOM is used to rapidly extinguish light incident on the cavity resulting in an exponential decay of the transmitted light with a $1/e$ time constant equal to $\tau=\mathcal{F}/(2\pi\nu_\mr{FSR})$ \cite{rempe92}. We record a time constant of $\tau~=~15.1(9)~\mu$s at 852~nm, resulting in a finesse $\mathcal{F}=1.42(8)\times10^5$ and cavity linewidth $\delta\nu=10.5$~kHz. As expected from the coating specification, at 1018~nm a reduced time constant of 4.1(4)~$\mu$s is measured, corresponding $\mathcal{F}=3.9(4)\times10^4$ and $\delta\nu=38.5$~kHz.

\begin{figure}[t]
\centering \includegraphics{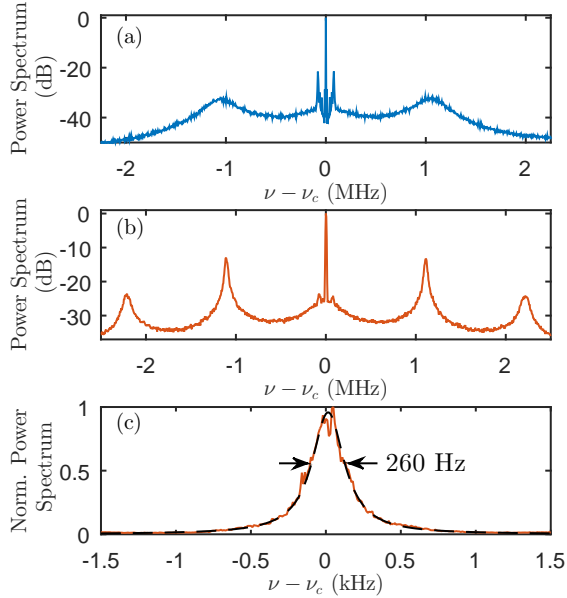}
\caption{(a) In-loop error signal for Rydberg B plotted relative to the PDH frequency with 5~kHz resolution bandwidth (RBW) showing servo bandwidth of 1.1~MHz. (b) Optical beatnote relative to $\nu_c=440$~MHz recorded at 509~nm between Rydberg A and B locked to consecutive $\mr{TEM_{00}}$ and TEM$_{01}$ modes with 10~kHz RBW. (c) Linearized power spectrum recorded with 10~Hz RBW showing Lorentzian linewidth FWHM=260(5)~Hz. Data represent 100 r.m.s. averages using 190~ms sweep time.}
\label{fig4}
\end{figure}

Direct measurement of laser linewidth for sub-kHz lasers is challenging and requires either multiple stable lasers, a narrow atomic reference or a sufficiently long optical fiber to perform delayed self-heterodyne interferometry \cite{okoshi80}. At 852~nm we are unable to perform either comparison, however using the two Rydberg lasers A \& B the linewidth can be measured from an optical beatnote at 509~nm. Rydberg A is locked to the TEM$_{00}$ mode of the ULE cavity while Rydberg B is locked to the TEM$_{01}$, with a frequency spacing of 220~MHz in the infra-red. Figure~\ref{fig4}(b) \& (c) show the optical beat note recorded on an rf spectrum analyzer with each trace the RMS average of 100 shots recorded with a 190~ms sweep time. Fig.~\ref{fig4}(b) reveals secondary peaks at harmonics of 1.1~MHz corresponding to the fast-feedback servo bumps for each laser. Fitting the central peak to a Lorentzian in Fig.~\ref{fig4}(c) returns a linewidth of 260(5)~Hz relative to the cavity, from which we can estimate a linewidth of $\sim130$~Hz for each laser due to the fact both lasers are locked using identical components. Whilst this measurement may underestimate the linewidth due to common mode noise rejection from locking to a single cavity, this results in an IR linewidth $<100$~Hz with better performance expected at 852~nm due to the increase in cavity finesse at this wavelength and an increased servo bandwidth of 1.2~MHz.

Converting the observed laser linewidth to gate fidelity is complex due to the error in a two-photon Raman transition being related to the relative phase noise between the two lasers \cite{saffman05a} which due to their different wavelengths cannot be measured without performing gate operations on a single qubit. As the lasers are locked to a common cavity the fluctuations are also correlated meaning the linewidths are not additive. However, using the available laser power an effective two-photon Rabi frequency of $\Omega/2\pi=10$~MHz can be achieved with a few GHz intermediate state detuning \cite{johnson08}. For a relative linewidth of 100~Hz, this results in an averaged gate error $\varepsilon\simeq10^{-6}$ when modelling the linewidth as a dephasing term following Ref. \cite{geabanacloche95}. Thus the laser system is suitable for high fidelity gates, with the laser-limited coherence time greatly exceeding the gate duration.

\section{Atomic Spectroscopy - Rydberg EIT }
\begin{figure}[t!]
\centering \includegraphics{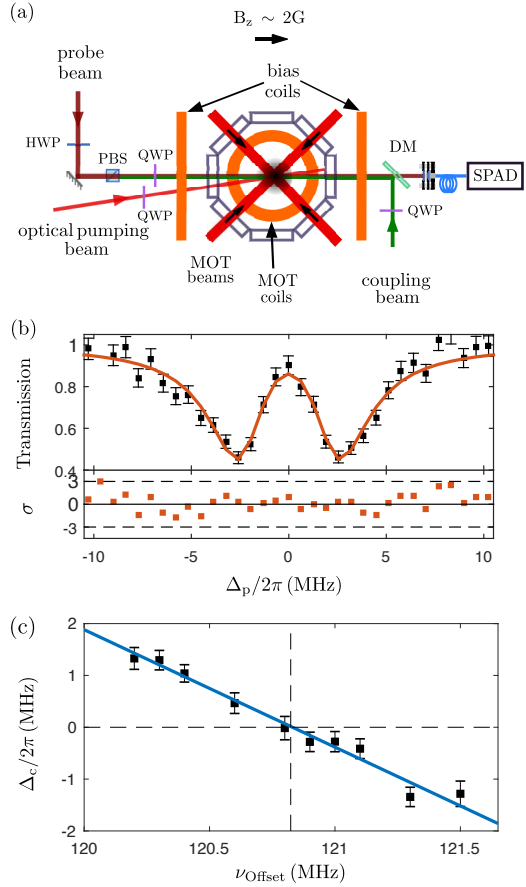}
\caption{(a) Schematic of the experimental EIT Setup. A strong coupling beam (green) counter-propagates with a weak probe beam (red) through a cold atom cloud of Cs atoms optically pumped in the $\ket{\mr{F \, = \, 4, m_{F} \, = \, 4}}$ dark state. (b) EIT transmission peak for 50S$_{1/2}$ Rydberg state with a coupling power of 50 mW with fitted $\chi^2_\nu=1.0$. Error bars represent one standard deviation. (c) Coupling laser detuning $\Delta_{c}/2 \pi$ as a function of the coupling laser offset frequency, fitted according to the formula $\Delta_\mr{c}/2 \pi=\alpha(\nu_\mr{Offset}-\nu_0)$, showing the two-photon resonance at $ \nu_{0} \, = \, 120.82(4)~\mr{MHz}$. }
\label{fig5}
\end{figure}

To calibrate ULE cavity mode frequencies with respect to Rydberg transitions we have performed high resolution spectroscopy on a cold atomic ensemble using Rydberg EIT \cite{mohapatra07,pritchard10}. Measurements are performed using the 50S$_{1/2}$ Rydberg state to reduce sensitivity to stray electric fields and minimize interaction effects. This is chosen as one of the lowest $n$ levels accessible with the current setup. Experiment are performed on a laser cooled $\mr{^{133}Cs}$ atom cloud with the experimental setup shown schematically in Fig.~\ref{fig5}(a). Atoms are loaded into a magneto-optical trap for 1~s followed by a short polarization gradient cooling stage resulting in 10$^5$ atoms at a temperature of $5~\mu$K. Atoms are then prepared in the $\ket{F=4, m_{F}=4}$ stretched state by dark-state optical pumping with $\sigma^{+}$ polarized light on the $6S_{1/2}~F=4\rightarrow6P_{3/2}~F'=4$ transition using a 2~G magnetic field along the probe beam to define a quantization axis.

EIT spectroscopy is performed using~counter-propagating probe and coupling  lasers focused to 1/e$^{2}$~radii of 13 $\mu$m and 84 $\mu$m, respectively. The probe laser is derived from the cooling laser, stabilized using polarization-spectroscopy on the 6s$^{2}$ S$\mr{_{1/2}}$ $\rightarrow$ 6p$^{2}$ P$\mr{_{3/2}}$ transition. The coupling laser, Rydberg B, is offset-locked to the cavity and drives the 6p$^{2}$ P$\mr{_{3/2}}$ $\rightarrow$ 50s$^{2}$ S$\mr{_{1/2}}$ with a power of 50~mW. The probe and coupling lasers are circularly polarized in a $\sigma^{+} \, - \, \sigma^{-}$ configuration with respect to the quantization axis, maximizing the transition amplitude to the Rydberg state. The probe power is fixed at 0.5 pW ($\mr{I/I_{sat} \, \sim 5 \, \times \, 10^{-\,4}}$) and the probe transmission is recorded using a single photon counting module (SPAD). For each measurement, the probe laser is scanned across the resonance from $\Delta_{p}/2\pi=-12 \rightarrow+12$~MHz in 1~ms, with spectra recorded from the average of 100 shots. Data are recorded as a function of the coupling laser offset frequency, $\nu_\mr{Offset}$, to find the resonant frequency for the upper transition with respect to the cavity resonance. 

Figure~\ref{fig5}(b) shows a typical EIT spectrum for a sideband frequency $\nu_\mr{Offset}=120.8$~MHz, which is fit to the weak-probe susceptibility \cite{geabanacloche95} enabling determination of the Rabi frequency, $\Omega_c/2\pi=5.1(2)$~MHz, and the coupling laser detuning $\Delta_{c}$. Error bars represent one standard deviation. The normalized residuals on the lower panel show excellent agreement between theory and experiment, with $\chi^2_\nu=1.0$. From the complete set of spectra, the coupling laser detuning as a function of sideband frequency is plotted in Fig.~\ref{fig5}(c) and fit using the formula $\Delta_\mr{c}=\alpha(\nu_\mr{Offset}-\nu_0)$, enabling determination of the resonant sideband frequency as $\nu_0=120.82(4)$~MHz with a precision of less than 0.1~MHz.

\begin{figure}
\centering \includegraphics{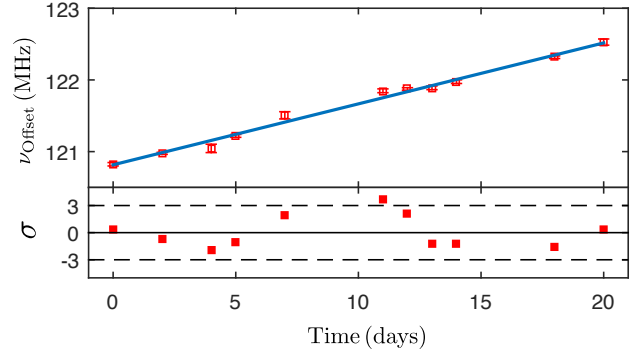}
\caption{EIT resonance's frequency of $\mr{50S_{1/2}}$ Rydberg state recorded for a period of 20 days. Using a linear fit, the constant cavity's drift is evaluated at $\mr{1 Hz/s}$. Inset show normalised residuals. Errorbars reflect standard errors.}
\label{fig6}
\end{figure}

Using this high resolution method to extract the resonant transition frequency, we repeat the EIT spectroscopy over a period of 20~days to determine the long-term frequency drift of the cavity with respect to the atomic transition as shown in Fig.~\ref{fig6}. The results show an average linear frequency drift of 1~Hz/s, confirming the ULE cavity is optimized close to the zero-CTE temperature but still limited by creep of the spacer or outgassing in the ULE cavity vacuum. 

\section{Conclusion}
We have demonstrated sub-kHz linewidth lasers for Rydberg excitation with three lasers locked simultaneously to the same high-finesse ULE reference cavity. The lock configuration allows continuous tuning of the laser offset using the "electronic sideband" technique, and after frequency doubling we measure laser linewidth of $\mr{\sim \, 130 \,Hz}$ at 509~nm suitable for performing high fidelity quantum gates. Using high-resolution EIT spectroscopy on a cold atom cloud, we calibrate cavity frequencies with respect to Rydberg transitions with a precision of $<0.1$~MHz. Finally, we have demonstrated excellent long-term stability with a linear drift of $ \mr{\sim  \,1 \, Hz/s} $ relative to an atomic reference. These measurement are competitive against doubly-stabilized optical clocks \cite{hill16, bridge16} and offer an order of magnitude improvement compared to similar cavity stabilized Rydberg laser systems \cite{arias17, dehond17}.\\

\subsection*{Acknowledgments}
We acknowledge Erling Riis for useful discussions and careful reading of the manuscript. This work is supported by the UK Engineering and Physical Sciences Research Council, Grant No. EP/N003527/1. The data presented in the paper are available for download\cite{legaie17data}.


\end{document}